\documentclass [12 pt]{article}
\def\be{\begin{equation}}
\def\eea{\end{eqnarray}}
\def\bea{\begin{eqnarray}}
\def\ee{\end{equation}}

\usepackage[psamsfonts]{amssymb}
\usepackage{amsmath}
\author{Fardin Kheirandish$^{1}$ \footnote{fardin$_{-}$kh@phys.ui.ac.ir} and
Morteza Soltani$^{1}$ \footnote{soltani@phys.ui.ac.ir}
\\ $^{1}$ {\small Department of Physics, University of Isfahan,}
\\ {\small Hezar Jerib Avenue, Isfahan, Iran.}}
\title{Extension of the Huttner-Barnett model to a magnetodielectric medium}
\begin{document}
\maketitle
\begin{abstract}
The Huttner$-$Barnett model is extended to a magnetodielectric
medium by adding a new matter field to this model. The
eigenoperators for the coupled system are calculated and
electromagnetic field is written in terms of these operators. The
electric and magnetic susceptibility of the medium are explicitly
derived and shown to satisfy the Kramers$-$Kronig relations. It is
shown that the results obtained in this model are equivalent to the
results obtained from the phenomenological methods.
\noindent.\\
{\bf PACS number(s): 12.20.Ds}
\end{abstract}
\section{Introduction}
Electromagnetic field (EM) is traditionally quantized by associating
a quantum-mechanical harmonic oscillator with each mode of the
radiation field in free space \cite{1,2}. This is achieved by
introducing a general position and momentum and expressing the modes
in terms of them. The transition from classical to quantum
mechanical description takes place by treating the generalized
position and momentum quantum mechanically and demanding equal-time
commutation relations (ETCR) between them. This scheme is called
canonical quantization and is based on a Lagrangian density.

In the presence of a linear polarizable medium canonical
quantization is retained if the Hattner-Barnett model (HB) is used
\cite{3}. This model is based on the Hopfield model \cite{4} and
the matter is modeled by a harmonic field which represents the
medium excitations. The inclusion of losses into the system can be
done by inserting a reservoir consisting of a continuum of
harmonic oscillators \cite{5,6}. Although this model is based on a
microscopic model of matter, any microscopic parameter like
coupling function between matter and EM field is not appeared in
final results and in fact the results are written in terms of the
electric susceptibility $\chi_e$ which is a macroscopic parameter
of the medium.

Since the Hopfield model is written for a homogeneous polarizable
matter and not for a magnetizable one, and since the HB model is
based on this model, so we should extend the Hopfield model to a
magnetizable medium. As mentioned, the final results of HB model do
not depend on microscopic parameters, so we can change the
microscopic relations such that we get the correct Maxwell equations
in the magnetizable medium.

Although by doing so, we may lose the physical interpretation of
microscopic interaction, but at least we have a Lagrangian density
from which correct Maxwell equations are obtained and a canonical
quantization scheme is introduced. In addition, this model has a
clear assumption about matter which can not be found in other
macroscopic methods.

The common method for the EM field quantization in the presence of
matter is the macroscopic method (phenomenological method). The
macroscopic approach to EM field quantization begins with Maxwell
equations and the loss is modeled by the Langevin force in the form
of the noise current operators. A correlation function proportional
to the imaginary part of the dielectric function is assumed for the
later operations. In this way, a straightforward calculation lead to
the field expression in terms of the noise operators and the Green
function of Maxwell equations \cite{7}$-$\cite{11}. In this method
ETCR between the EM field operators are postulated and the
verification of them justifies the validity of this approach. This
model is recently extended to a magnetodielectric medium by adding a
new noise operator for magnetization and assuming a correlation
between them which is proportional to the imaginary part of magnetic
susceptibility \cite{12}. In the conclusion of this paper, we will
make a comparison between the results of our model and the
macroscopic quantization model.

The paper is arranged as follows: In Sec. 2, the basic theory of
electrodynamics in a magnetodielectric medium is reviewed. In sec.
3, we extend the HB model to a magnetizable medium which is a
special case of a magnetodielectric medium, in fact, this section is
an introduction to the more generalized case of a magnetodielectric
medium. In sec. 4, we obtain a Lagrangian density in a
magnetodielectric medium from which we quantize the EM field.
Finally, the main points are summarized in Sec. 5.
\section{Basic equations}
The classical Maxwell equations in a magnetodielectric medium are
\begin{equation}\label{1}
\nabla  \times {\bf E}({\bf r},t) + \frac{{\partial {\bf B}({\bf
r},t)}}{{\partial t}} = 0,
\end{equation}

\begin{equation}\label{2}
\nabla \times{\bf H}({\bf r},t)-\frac{{\partial {\bf D}({\bf
r},t)}}{{\partial t}} = 0.
\end{equation}
The constitutive relations relate ${\bf D}$ and ${\bf H}$ to ${\bf
E}$ and ${\bf B}$. For isotropic linear medium, the relation
between polarization and electric field is
\begin{equation}\label{3}
{\bf P}({\bf r},t) = \varepsilon _0 \int_{ - \infty }^t dt' \chi _e
({\bf r},t - t')\,{\bf E}({\bf r},t')\,dt',
\end{equation}
where $\varepsilon_0$ is the permittivity of free space. The
magnetization and the magnetic induction field are related via a
similar expression,
\begin{equation}\label{4}
{\bf M}({\bf r},t) = \kappa _0 \int_{ - \infty }^t dt' \chi _m ({\bf
r},t - t')\,{\bf B}({\bf r},t')\,dt',
\end{equation}
where $\kappa_0$ is the inverse permeability of free space. The
electric and magnetic susceptibilities are defined in the frequency
domain as
\begin{equation}\label{5}
\chi _\sigma  ({\bf r},\omega ) = \int_0^{ + \infty }\,d\tau\,\chi
_\sigma  ({\bf r},\tau )\,e^{\imath\omega \tau },
\end{equation}
where the index $\sigma$ could be $e$ or $m$. The electric
displacement ${\bf D}$ is expressed in terms of the electric field
${\bf E}$ and the polarization field ${\bf P}$ as
\begin{equation}\label{6}
{\bf D}({\bf r},t) = \varepsilon _0 {\bf E}({\bf r},t) + {\bf
P}({\bf r},t).
\end{equation}

The constitutive relation which defines the magnetic field ${\bf
H}$ in terms of the magnetic induction ${\bf B}$ and the
magnetization ${\bf M}$ is

\begin{equation}\label{7}
{\bf H}({\bf r},t) = \kappa _0 {\bf B}({\bf r},t) - {\bf M}({\bf
r},t).
\end{equation}

The real and imaginary parts of the electric and magnetic
susceptibilities should satisfy Kramers-Kronig relations. This shows
that the dissipative nature of medium is an immediate consequence of
its dispersive character and vice versa. Therefore, in general we
are dealing with a dissipative problem. For investigating a
dissipative quantum system we should include a reservoir in the
process of quantization. After the quantization, the definition of
polarization ${\bf P}$, and magnetization ${\bf M}$, in terms of the
Fourier components should be reformed as
\begin{equation}\label{8}
\hat{{\bf P}}({\bf r},t) = \varepsilon _0 \int_{0}^\infty
d\omega\{[\chi _e ({\bf r},\omega)\hat{{\bf E}}({\bf
r},\omega)+\hat{{\bf P}}_N({\bf r},\omega)]e^{\imath\omega
t}+H.c.\},
\end{equation}
and
\begin{equation}\label{9}
\hat{{\bf M}}({\bf r},t) = \kappa _0 \int_{0}^{\infty}d\omega\{[
\chi _m ({\bf r},\omega)\hat{{\bf B}}({\bf r},\omega)+\hat{{\bf
M}}_N({\bf r},\omega)]e^{\imath\omega t}+H.c.\},
\end{equation}
respectively, where ${\bf \hat{P}}_N({\bf r},\omega)$ and ${\bf\hat{
M}}_N({\bf r},\omega)$ are polarization
 and magnetization noise operators and satisfy the fluctuation-dissipation theorem \cite{13}.
\section{ Extension of the Huttner-Barnet model
to a magnetizable medium} Here, knowing the Lagrangian density of HB
model, we change the form of the interaction term and find a
Lagrangian density in a magnetizable medium. Let us write the
Lagrangian as
\begin{equation}\label{10} {\cal L} = {\cal
L}_{em}  + {\cal L}_{mat}  + {\cal L}_{res}  + {\cal L}_{int},
\end{equation}
where
\begin{equation}\label{11}
 {\cal L}_{em}  = \frac{{\varepsilon _0 }}{2}{\bf E}^2  -
\frac{\kappa_0}{{2}}{\bf B}^2,
\end{equation}
is the EM part which can be expressed in terms of vector potential
$\textbf{A}$ and scaler potential $U$ (${\bf E}=-\dot{\bf A}-\nabla
U $ and $\textbf{B}=\nabla\times \textbf{A}$).

\begin{equation}\label{12}
{\cal L}_{mat}= \frac{\rho }{2}{\bf \dot X}^2- \frac{{\rho \omega
_0^2}}{2}{\bf {X}}^2,
\end{equation}
is the matter part, modeled by a harmonic oscillator field ${\bf X}$
of frequency $\omega_0$.
\begin{equation}\label{13}
{\cal L}_{res}  = \int_0^\infty  {d\omega } \left( {\frac{\rho
}{2}{\bf \dot Y}_\omega ^2  - \frac{{\rho \omega ^2 }}{2}{\bf
Y}_\omega ^2 } \right)
\end{equation}
is the reservoir part, consisting of a continuum of harmonic
oscillators and
\begin{equation}\label{14}
{\cal L}_{int}= \alpha \nabla \times{\bf A}\cdot{\bf X} -
\int_0^\infty{d\omega\,\nu (\omega )\,{\bf X}\cdot\dot{\bf
Y}_\omega},
\end{equation}
is the interaction part which includes the interaction between the
light and the magnetization field, with coupling constant $\alpha$,
and also the interaction between the magnetization field and the
reservoir oscillator field with the frequency dependent coupling
constant $\nu(\omega)$. In this equation the interaction term
between the EM and matter field is chosen such that it lead to
correct Maxwell equations in a magnetizable medium.

We make two assumptions about $\nu(\omega)$: (i) the analytic
continuation of $\mid\nu(\omega)\mid^2$ to negative frequency is an
even function and (ii) $\nu(\omega)\neq0$ for all nonzero
frequencies. The first assumption is needed in order to extend the
 frequency integrals to the negative real axis, while the second
one ensure that all the reservoir oscillators couple to the system.

 For a purely magnetizable medium $\chi_e=0$, and we can make $U=0$ by
 choosing the Coulomb gauge ($\nabla\cdot{\bf A}=0$). Let us define ${\bf
M}=\alpha{\bf X}$ and write ${\bf H}$ as

\begin{equation}\label{15}
{\bf H}=\kappa_0{{\bf B}}-{\bf M}.
\end{equation}
Later we prove that $\bf H$ in Eq.(\ref{15}) is in fact the same
$\bf H$ which appears in Maxwell equations.

Now we go to reciprocal space and write all the fields in terms of
their spatial Fourier transforms. For example electric field is
written as
\begin{equation}\label{16}
{\bf E}({{\bf
r},t})=\frac{1}{{({2\pi})^{\frac{3}{2}}}}\int{d^3\,{\bf k}}\,{\bf
\tilde E}({\bf k},t)\,e^{\imath{\bf k}\cdot{\bf r}}
\end{equation}
(tilde distinguishes the field in real and reciprocal spaces). The
total Lagrangian in reciprocal space is
\begin{equation}\label{17}
L = \int'{d^3{\bf k}\,(\tilde{\cal L}_{em}+\tilde{\cal L}_{mat}+
\tilde {\cal L}_{res}+\tilde{\cal L}_{int})}.
\end{equation}
Since
$\tilde{\textbf{F}}^*(\textbf{k},t)=\tilde{\textbf{F}}(-\textbf{k},t)$
is satisfied for any arbitrary real field we should restrict the
integration to half of the reciprocal space, which is indicated by a
prim over the integral in (\ref{17}). The Lagrangian densities in
the reciprocal space are obtained as
\begin{equation}\label{18}
\tilde{{\cal L}}_{em}=\varepsilon_0({\bf \tilde E}^2-c^2{\bf
\tilde B}^2),
\end{equation}

\begin{equation}\label{19}
\tilde{{\cal L}}_{mat}=\rho \dot{\tilde{\textbf{X}}}^{2}-\rho
\omega _0^2{\bf \tilde X}^{2},
\end{equation}

\begin{equation}\label{20}
\tilde{{\cal L}}_{res}= \int_0^\infty {d\omega }\, (\rho{\dot
{\tilde{\textbf{Y}}}}_\omega ^{2}  - {\rho \omega ^2 }\tilde{{\bf
Y}}_\omega ^{2}),
\end{equation}

\begin{equation}\label{21}
\tilde{\cal L}_{int}=\alpha
\textbf{k}\times\tilde{\bf{A}}^*\cdot\tilde{\bf{X}}-
\int_0^\infty{d\omega\nu(\omega
)\tilde{\bf{X}}\cdot\dot{\tilde{\textbf{Y}}}_\omega^{*}+c.c..}
\end{equation}
The Coulomb gauge in reciprocal space can be written as
$\textbf{k}\cdot\tilde{\textbf{A}}(\textbf{k},t)=0$. Using the
definition of longitudinal part of matter field, that is
$\textbf{k}\times\tilde{\textbf{X}}^\parallel(\textbf{k},t)=0$, and
the coulomb gauge condition, it is easy to prove that ${\bf
k}\times\tilde{\textbf{A}}(\textbf{k},t)\cdot\tilde{\textbf{X}}^\parallel
(\textbf{k},t)=0$. This shows that there is no interaction between
the longitudinal part of the matter field and EM field. Therefor
without losing any generality we can consider matter and reservoir
fields as transverse fields.

 We introduce unit polarization vectors ${\bf e}_\lambda({\bf
k})$, $\lambda=1,2$, which are orthogonal to $\hat{\textbf{k}}$ and
to each other, and decompose the transverse fields along them to get
\begin{equation}\label{22}
\tilde{{\bf A}}({\bf k},t) = \sum\limits_{\lambda  = 1,2}
{\tilde{A}_\lambda } ({\bf k},t)\,{\bf e}_\lambda  ({\bf k})
\end{equation}
with similar expressions for the other fields. $\tilde{{\cal L}}$
can be used to obtain the conjugate components of the fields

\begin{equation}\label{23}
-\varepsilon_0\tilde{E}_\lambda=\frac{\partial{\cal
L}}{\partial\dot{\tilde{A}}_\lambda^*}=\varepsilon_0\dot{\tilde{A}}_\lambda,
\end{equation}

\begin{equation}\label{24}
\tilde {P}_{\lambda}=\frac{\partial{\cal L}}
{\partial{\dot{\tilde{X}}_\lambda^*}}=
\rho\dot{\tilde{X}}_{\lambda},
\end{equation}

\begin{equation}\label{25}
\tilde{Q}_{\omega\lambda}= \frac{\partial{\cal L}}{\partial
\dot{\tilde{Y}}_{\omega\lambda}^*}=\rho\dot
{\tilde{Y}}_{\omega\lambda}-v(\omega)\tilde{Y}_{\omega\lambda}.
\end{equation}

Using the Lagrangian (\ref{17}) and the expressions for the
conjugate variables (\ref{23})-(\ref{25}), we obtain the Hamiltonian
as
\begin{equation}\label{26}
 H=\int' d^3\textbf{k}(\tilde{\cal H}_{em}+\tilde{\cal
H}_{mat}+\tilde{\cal H}_{int}),
\end{equation}
where

\begin{equation}\label{27}
\tilde{\cal H}_{em}=\varepsilon_0(\tilde{\textbf{E}})^2
+\varepsilon_0c^2\textbf{k}^2\tilde{\textbf{A}}^2.
\end{equation}
and

\begin{eqnarray}\label{28}
\tilde{\cal H} _{mat}  &=& \frac{\tilde{\textbf{P}}^{2}}{\rho
}\mathop{+\rho\tilde{\omega}_0^2\tilde{\textbf{X}}^{2}+}\int_0^\infty
d\omega (\frac{{\tilde{\textbf{Q}}_\omega}}{{\rho }}^2+{\rho} \omega
^2 \tilde{\textbf{Y}}_\omega ^{2})\nonumber\\&+&\int_0^\infty
{d\omega } \frac{{(\nu(\omega)}}
{\rho}\tilde{\textbf{X}}^{*}\cdot\tilde{\textbf{Q}} _\omega) + c.c.,
\end{eqnarray}
is the matter part, including the interaction between the
magnetization and the reservoir.
 $\tilde{\omega}_0^2\equiv\omega_0^2+\int_0^\infty
d\omega\frac{v(\omega)^2}{\rho^2}$ is the renormalized frequency of
the polarization field and

\begin{equation}\label{29}
H_{int}=-\alpha\int{d^3{\bf k}}[{\bf k}\times\tilde{{\bf
A}}^*\cdot\tilde{{\bf X}}+ c.c.],
\end{equation}
is the interaction between the EM field and the magnetization. The
fields are quantized in a standard fashion by demanding ETCR between
the variables and their conjugates. For EM field components we have

\begin{equation}\label{30}
[\hat{\tilde{A}}_\lambda(\textbf{k},t),\hat{\tilde{E}}_{\lambda'}^{{*}}
(\textbf{k}',t)]=\frac{-\imath\hbar}{\varepsilon_0}\,\delta
_{\lambda\lambda'}\,\delta(\textbf{k}-\textbf{k}'),
\end{equation}
and for the matter fields

\begin{equation}\label{31}
[\hat{\tilde{X}}_{\lambda}(\textbf{k},t),\hat{\tilde{P}}_{\lambda'}^{*}(\textbf{k}',t]
= \imath\hbar\,\delta_{\lambda,\lambda'}\,\delta (\textbf{k}-
\textbf{k}'),
\end{equation}

\begin{equation}\label{32}
[\hat{\tilde{Y}}_{\omega\lambda}(\textbf{k},t),\hat{\tilde{Q}}
_{\omega'\lambda'}^{{*}}(\textbf{k}',t)]=\imath\hbar\,\delta
_{\lambda\lambda'}\,\delta(\textbf{k}-\textbf{k}')\,\delta(\omega-\omega'),
\end{equation}
with all other equal-time commutators being zero.

Using the inverse Fourier transform, the obtained Hamiltonian in
(\ref{26}) could be written in coordinate space as
\begin{eqnarray}\label{33}
\hat{H}&=&\int {d^3 {\bf r}} [\frac{{\varepsilon _0 \hat{{\bf
E}}^2(\textbf{r},t) }}{2} + \frac{{(\kappa_0\nabla  \times \hat{{\bf
A}}(\textbf{r},t))^2 }}{{2}} - \nabla \times \hat{{\bf
A}}(\textbf{r},t)\cdot\hat{{\bf M}}(\textbf{r},t)]\nonumber\\&+&
\hat{H}_{mat}.
\end{eqnarray}
The commutation relation between  $\hat{\textbf{E}}(\textbf{r},t)$
and $\hat{\textbf{A}}(\textbf{r},t)$ from, (\ref{30}), is obtained
as

\begin{equation}\label{34}
 [\hat{\textbf{A}}_\lambda({\textbf{r},t}),\hat{\textbf{E}}
 _{\lambda'}(\textbf{r}',t)]=
 \imath\frac{\hbar}{\varepsilon_0}\delta_{\lambda\lambda'}^\perp
({\textbf{r}-\textbf{r}'}),
\end{equation}
where $\delta^\perp_{\lambda\lambda'}({\bf r}-{\bf r'})$ is the
transverse delta function \cite{14}. Using the commutation
relations (\ref{34}) and the Hamiltonian (\ref{33}), we find the
Heisenberg equations for $\bf{A}({\textbf{r},t})$ and
$\bf{E}({\textbf{r},t})$ as

\begin{equation}\label{35}
\hat  {\bf E}(\textbf{r},t) = \frac{\partial  \hat{\bf
A}(\textbf{r},t)}{\partial t},
\end{equation}
and
\begin{equation}\label{36}
\varepsilon_0\frac{{\partial \hat{\bf E}(\textbf{r},t)}}{{\partial
t}} = \kappa_0\nabla \times \nabla \times \hat{{\bf
A}}(\textbf{r},t)-\nabla \times \hat{{\bf M}}(\textbf{r},t).
\end{equation}
Eq.(\ref{35}) is one of the Maxwell equations and Eq.(\ref{36}) is
also the Maxwell equation in a magnetizable medium if we accept
relation (\ref{15}) and interpret $\hat{\bf M}(\textbf{r},t)$ as the
magnetization. This shows that the assumed interaction between EM
field and matter field lead to the proper Maxwell equations in a
magnetic medium.

In the following instead of solving the Heisenberg equation for EM
field we write the EM field in terms of eigenoperators of the
Hamiltonian (\ref{26}).

To facilitate the calculations, we introduce a set of three
annihilation operators as

\begin{equation}\label{37}
\hat{a}_\lambda(\textbf{k},t) = \sqrt {\frac{\epsilon_0}{{2\hbar
kc}}}[kc \hat{\tilde{A}}_\lambda(\textbf{k},t)-\imath\hat
{\tilde{E}}_{\lambda}(\textbf{k},t)],
\end{equation}

\begin{equation}\label{38}
\hat{b}_\lambda(\textbf{k},t) = \sqrt {\frac{\rho }{{2\hbar
\tilde{\omega}_0}}}[\tilde{\omega}_0
\hat{\tilde{X}}_\lambda(\textbf{k},t)+\frac{\imath}{\rho}\hat
{\tilde{P}}_{\lambda}(\textbf{k},t)],
\end{equation}

\begin{equation}\label{39}
\hat{b}_{\lambda}(\textbf{k},\omega,t)=\sqrt{\frac{\rho}{{2\hbar
\omega}}}
[-\imath\omega\hat{\tilde{Y}}_{\omega\lambda}(\textbf{k},t)+
\frac{1}{\rho}\hat{\tilde{Q}}_{\omega\lambda}^{*}(\textbf{k},t)].
\end{equation}
From the ETCR for the fields, (\ref{30})-(\ref{32}), we obtain the
ETCR for above operators as

\begin{equation}\label{40}
[\hat{a}_{\lambda}(\textbf{k},t),
\hat{a}_{\lambda}^\dag(\textbf{k}',t)]=\delta_{\lambda
\lambda'}\delta(\textbf{k}-\textbf{k}'),
\end{equation}

\begin{equation}\label{41}
[\hat{b}_{\lambda}(\textbf{k},t),\hat{b}_{\lambda}^\dag(\textbf{k}',t)]=\delta_{\lambda
\lambda'}\delta(\textbf{k}-\textbf{k}'),
\end{equation}

\begin{equation}\label{42}
[\hat{b}_{\lambda}(\textbf{k},\omega,t),\hat{b}_{\lambda'}^\dag(
\textbf{k}',\omega',t)]=\delta_{\lambda\lambda'}
\delta(\textbf{k}-\textbf{k}')\delta(\omega-\omega').
\end{equation}
We emphasize that, in contrast to the previous ETCR between
conjugate fields, which where correct only in half space,
Eqs.(\ref{40})-(\ref{42}) are valid in the hole reciprocal space. By
inverting (\ref{37})-(\ref{39}) to find the field operators, and
inserting these fields into the Hamiltonian(\ref{26}), we obtain
after integration
\begin{equation}\label{43}
\hat{H}_{em}=\int{d^3{\bf k}\sum\limits_{\lambda  = 1,2}} \hbar
\omega _{\bf k}\, \hat a_\lambda ^\dag \, ({\bf k})\,\hat a_\lambda
({\bf k}),
\end{equation}

\begin{eqnarray}\label{44}
\hat{H}_{mat}&=&\int d^3\textbf{k}\sum\limits_{\lambda=1,2}
\{\hbar\tilde{\omega}_0\,\hat{b}_\lambda^\dagger(\textbf{k})
\hat{b}_\lambda\,(\textbf{k})+\int_o^\infty
\omega\hbar\omega\,\hat{b}_{\lambda}^\dagger\,(
\textbf{k},\omega)\,\hat{b}_{\lambda}(\textbf{k},\omega)\nonumber\\
&+&\frac{\hbar}{2}\int_0^\infty
 d\omega\,
 V(\omega)\,[\hat{b}^\dagger_\lambda(-\textbf{k})+\hat{b}_\lambda
 (\textbf{k})][\hat{b}_{\lambda}^\dagger(-\textbf{k},\omega)
 +\hat{b}_{\lambda}(\textbf{k},\omega)]\},\nonumber\\
\end{eqnarray}

\begin{equation}\label{45}
\hat{H}_{int}  = \int {d^3 {\bf k}}\sum_{\lambda, \lambda'=1,2}
\alpha\sqrt {\frac{{\hbar \omega_\textbf{k}}}{{2\varepsilon _0 }}}
(\hat a_\lambda (\textbf{k}) + \hat a_\lambda ^\dag  ( -
\textbf{k}))\sqrt {\frac{\hbar }{{2\rho \tilde \omega _0 }}} (\hat
b_{\lambda '} (\textbf{k}) + \hat b_{\lambda '} ^\dag  ( -
\textbf{k}))\epsilon_{\lambda\lambda'},
\end{equation}
where $\epsilon_{\lambda\lambda'}$ is the antisymmetric symbol,
$\omega_\textbf{k}=c|\textbf{k}|$ and
$V(\omega)=[\frac{\nu(\omega)}{\rho}]\sqrt{\frac{\omega}{\tilde{\omega}}}$.

The polarization and reservoir parts of the Hamiltonian
$\hat{H}_{mat}$ (\ref{45}) can be diagonalized by using the Fano
technique \cite{16} to get a
 dressed matter field. The diagonalized expression for $\hat{H}_{mat}$ is (the detail of this
technique can be found in \cite{5} and we only give the results)
\begin{equation}\label{46}
\hat{H}_{mat}=\int_0^\infty{d\omega}
  \int{d^3\textbf{k}\sum\limits_{\lambda=1,2}
{\hbar\omega}\,\hat{B}_\lambda^\dag
  (\textbf{k},\omega)\,\hat{B}_\lambda(\textbf{k},\omega)},
 \end{equation}
where $\hat{B}^\dag(\textbf{k},\omega)$ and
$\hat{B}(\textbf{k},\omega)$ are creation and annihilation operators
of the dressed matter field respectively, which satisfy the usual
ETCR

\begin{equation}\label{47}
[\hat{B}_\lambda({\bf k},\omega),\hat{B}_\lambda'({\bf
k}',\omega')] =\delta_{\lambda\lambda'}\delta({\bf k}-{\bf
k}')\delta(\omega-\omega').
\end{equation}

They can be expressed in terms of the initial creation and and
annihilation operators as

\begin{equation}\label{48}
\hat{B}({\bf k},\omega ) = \alpha _0 (\omega )b({\bf k}) + \beta _0
(\omega )b^\dag  ({\bf k}) + \int_0^\infty \, {d\omega }\, \alpha
(\omega ,\omega ')\,b({\bf k},\omega ) + \beta (\omega ,\omega
')\,b^\dag  ({\bf k},\omega ),
\end{equation}
and all the coefficient $\alpha_0(\omega)$,
 $\beta_0(\omega)$, $\alpha_1(\omega,\omega')$
 and $\beta_1(\omega,\omega')$ can be obtained in terms of microscopic parameters.

Using the commutators of $\hat{b}$ with $\hat{B}$ and $\hat{B}^\dag$
together with (\ref{48}), it is easy to show that
\begin{equation}\label{49}
\hat{b}_\lambda(\textbf{k})=\int_0^\infty{d\omega[\alpha^*_0
(\omega)}\hat{B}_\lambda^\dag(\textbf{k},\omega)-\beta_0(\omega)
\hat{B}_\lambda(\textbf{k}\omega)].
\end{equation}

We call this model the damped magnetization model. In this model the
explicit form of $\alpha_0(\omega)$ and $\beta_0(\omega)$ in terms
of the microscopic parameter such as matter field density, $\rho$,
or the coupling between matter and reservoir, $\nu(\omega)$, is not
important and we can only accept that such a function exist.

In terms of the new set of operators the interaction part of the
Hamiltonian can be written as

\begin{equation}\label{50}
H_{int}= -\int {d^3 {\bf k}\int_0^\infty  {d\omega \sum_{\lambda,
\lambda'}\frac{\hbar }{2}\Lambda (k)g( \omega)B_\lambda ^\dag ({\bf
k},\omega)(\hat a_\lambda  ({\bf k}) + \hat a_\lambda ^\dag ( - {\bf
k})} } )\epsilon_{\lambda\lambda'}
\end{equation}
where $ \Lambda (k) = \sqrt
{\frac{\alpha^2\omega_\textbf{k}}{c^2\varepsilon _0 \rho \omega
_0}}$ and $g(\omega)=(\alpha^*_0(\omega)-\beta^*_0(\omega))$.

Now we can follow two different methods to obtain the time
dependence of EM field. One is writing the Heisenberg equation and
solving it by using the Laplace transformation. The second one is
using the Fano technique and finding the diagonalized $\hat{H}$ and
writing EM field in terms of new operators.

Here we following the second method and write the Hamiltonian as

\begin{equation}\label{51}
\hat{H}=\int_0^\infty{d\omega}
  \int{d^3\textbf{k}\sum\limits_{\lambda=1,2}
{\hbar\omega}\hat{C}_\lambda^\dag
  (\textbf{k},\omega)\hat{C}_\lambda(\textbf{k},\omega)},
 \end{equation}
where

\begin{eqnarray}\label{52}
\hat{C}_\lambda({\bf k},\omega ) &=& \tilde{\alpha} _0 ({k},\omega
)a_\lambda({\bf k}) + \tilde{\beta} _0 ({k},\omega )a_\lambda^\dag
({\bf k})\nonumber\\&+&\int_0^\infty {d\omega'}\sum_{\lambda'=1,2}
[\tilde{\alpha} ({k},\omega,\omega')B_{\lambda'}({\bf
k},\omega,\omega' ) + \tilde{\beta} ({k},\omega,\omega'
)B_{\lambda'}^\dag ({\bf k},\omega'
)]\epsilon_{\lambda\lambda'}.\nonumber\\
 \end{eqnarray}

If we follow the algebra of reference \cite{5} the coefficients in
the relation above, (\ref{52}), can be written as
\begin{equation}\label{53}
\tilde{\alpha}_0({k},\omega )=(\frac{{\omega+\omega_{\bf k}
}}{2})\frac{{V(\omega,{k})}}{{\omega^2-\omega_{\bf
k}^2z(\omega)}},
\end{equation}

\begin{equation}\label{54}
\tilde{\beta}_0({k},\omega )=(\frac{{\omega-\omega_{\bf k}
}}{2})\frac{{V(\omega,{k})}}{{\omega ^2  - \omega _{\bf
k}^2z(\omega) }},
\end{equation}

\begin{equation}\label{55}
\tilde{\alpha} ({k},\omega ,\omega ') = \delta (\omega  - \omega ')
+ (\frac{{\omega _{\bf k} }}{2})(\frac{{V^* (\omega',{k})}}{{\omega
- \omega ' - \imath\varepsilon }})(\frac{{V(\omega,{k})}}{{\omega ^2
- \omega _{\bf k}^2 }}),
\end{equation}

\begin{equation}\label{56}
\tilde{\beta}({k},\omega ,\omega ') = (\frac{{\omega _{\bf k}
}}{2})(\frac{V(\omega',{k})}{\omega+\omega'})(\frac{{V(\omega,{k}
)}}{{\omega ^2- \omega _{\bf k}^2 z(\omega )}}).
\end{equation}
where $V(\omega,k)=\Lambda(k)g(\omega)$ and
$z(\omega)=1-\frac{1}{2\omega_k}\int_{-\infty}^{+\infty}d\omega'\frac{|V(\omega',{\bf
k})|^2}{\omega'-\omega+i\epsilon}$ and $\epsilon\to 0^+$.

Following the method used in (\ref{49}) to derive the matter field
operator $\hat{b}$ in terms of the dressed matter operators
$\hat{B}$ and $\hat{B}^\dag$, we invert (\ref{52}) to write the
photon annihilation operators $\hat{a}$ and dressed matter operator
$\hat{B}$ in terms of the new operators $\hat{C}$ and $\hat{C}^\dag$
as
\begin{equation}\label{57}
\hat{a}_\lambda({\bf k})=\int_0^\infty
{d\omega}[\tilde{\alpha}_0^*({k},\omega)\hat{C}_\lambda({\bf
k},\omega )-\tilde{\beta} _0 ({k},\omega )\hat{C}_\lambda^\dag
({\bf k},\omega )],
\end{equation}

\begin{equation}\label{58}
\hat{B}_\lambda({\bf k},\omega ) = \int_0^\infty  {d\omega '}
[\tilde{\alpha} ^* ({k},\omega ',\omega )\hat{C}_\lambda({\bf
k},\omega ) - \tilde{\beta} ({k},\omega ',\omega
)\hat{C}_\lambda^\dag ({\bf k},\omega )].
\end{equation}

From Eq.(\ref{37}), $\hat{\textbf{A}}(\textbf{r},t)$ is given by
 \begin{equation}\label{59}
\hat{\textbf{A}}(\textbf{r},t)=\frac{1}{(2\pi)^{\frac{3} {2}}} \int
d^3\textbf{k}\sum_{\lambda=1,2}\sqrt{\frac{\hbar}{{2\epsilon_0\omega_\textbf{k}}}}
[\hat{a}_\lambda(\textbf{k},t)e^{\imath\textbf{k}
\cdot\textbf{r}}+H.c.]\textbf{e}_\lambda(\textbf{k}),
\end{equation}
and we can use this expression and relation (\ref{57}) to obtain
\begin{eqnarray}\label{60}
\hat{\textbf{A}}(\textbf{r},t)&=&\frac{1}{{(2\pi
)^{\frac{3}{2}}}}\int {d^3 {\bf
k}}\sum\limits_{\lambda=1,2}\sqrt{\frac{\hbar \omega^2 _{\bf
k}}{2\varepsilon _0}} \int_0^\infty  {d\omega } \frac{{f(\omega
)}}{{\omega ^2-\omega_{\bf k}^2+\omega_{\bf
k}^2\chi_m(\omega)}}\nonumber\\&\times&[\hat C_\lambda({\bf
k},\omega )e^{\imath(\textbf{k}\cdot \textbf{r}-\omega t)} +
H.c.]\textbf{e}_\lambda(\textbf{k})
\end{eqnarray}
 where $f(\omega)$ is defined as $f(\omega)=\frac{\alpha g(\omega)}
 {\sqrt{c^2\varepsilon_0\rho\omega_0}}$ and $\chi_m(\omega)$ is
 defined by
\begin{eqnarray}\label{61}
\chi_m(\omega)=\frac{1}{2}\int_{-\infty}^{+\infty}{d\omega'}
\frac{{\left|{f(\omega')}\right|^2}}{{\omega'-\omega-
\imath\varepsilon}}= \frac{1}{2}P\int_{ - \infty }^{ + \infty } {}
\{ \frac{{|f(\omega ')|^2 }}{{\omega ' - \omega }}\} d\omega ' +
\frac{1}{2}\imath\pi
|f(\omega )|^2.\nonumber\\
\end{eqnarray}
Later we will show that $\chi_m(\omega)$ in Eq.(\ref{61}) is the
magnetic susceptibility.
 It can be seen from (\ref{61}) that the obtained magnetic susceptibility
 satisfies the Kramers$-$Kronig relations. In addition $f(\omega)$ can
 be written as

\begin{equation}\label{62}
|f(\omega)|^2=\frac{2Im\chi_m(\omega)}{\pi}.
\end{equation}

Let us write (\ref{60}) as

\begin{eqnarray}\label{63}
\hat{\textbf{A}}(\textbf{r},t)&=&\imath(\frac{\hbar}{{8\pi^4 }})\int
{d^3 {\bf k}}\sum\limits_{\lambda=1,2}\int_0^\infty {d\omega }
\frac{\omega_{\bf k}\sqrt{Im\chi_{m}(\omega)}}{{\omega
^2-\omega_{\bf k}^2(1-\chi_m(\omega))}}\nonumber\\&\times&[\hat
C_\lambda({\bf k},\omega )e^{\imath(\textbf{k}\cdot
\textbf{r}-\omega t)} - H.c.]\textbf{e}_\lambda(\textbf{k}).
\end{eqnarray}
Relation (\ref{63}) only depends on magnetic susceptibility which is
a macroscopic quantity. This result is exactly the same as the
result of macroscopic approach in a magnetizable medium \cite{12}.

Now we show that $\chi_m(\omega)$ is the Fourier transform of the
magnetizability of the medium. For this purpose we write
\begin{equation}\label{64}
\hat{{\bf X}}(\textbf{r},t) = \sqrt {\frac{\hbar }{2\rho\omega _0}}
\int \,d^3 {\bf k}\,\sum_{\lambda=1,2}\,[\hat{b}_\lambda ({\bf k})\,
e^{\imath {\bf k}\cdot {\bf r} } +
H.c.]\,\textbf{e}_{\lambda}(\textbf{k}),
\end{equation}
and using relations (\ref{49}) and (\ref{58}) we easily obtain
\begin{eqnarray}\label{65}
\hat{\tilde{M}}_\lambda(\textbf{k},\omega)&=&\sum_{\lambda'}
\{\sqrt{\frac{\hbar}{2\rho\omega_0}}\int_{0}^{+\infty } {d\omega
'[g(\omega )\alpha (\textbf{k},\omega ',\omega )+g^* (\omega )\beta
(\textbf{k},\omega ',\omega )]}
\nonumber\\&\times&\hat{C}_{\lambda'}(\textbf{k},\omega
')\epsilon_{\lambda'\lambda}\}.
\end{eqnarray}
Using (\ref{60}), Eq.(\ref{65}) can be written in real space as
\begin{equation}\label{66}
\hat{{\bf M}}({\bf r},t)=\int_0^\infty
d\omega\{[\kappa_0{\chi_m(\omega)} \nabla\times
\hat{\textbf{{A}}}({\bf r},\omega)+\hat{{\bf
M}}_N(\textbf{{r}},\omega)] e^{-\imath\omega t}+H.c.\}.
\end {equation}

Comparing (\ref{66}) with constitutive equation in (\ref{8}), the
interpretation of $\chi(\omega)$ as a magnetic susceptibility is
confirmed. In addition, the magnetization noise operator in
(\ref{8}) is obtained as

\begin{equation}\label{67}
\hat{\textbf{M}}_N({\bf r},\omega)=\int d^3
\textbf{{k}}\sum_{\lambda'=1,2}\sqrt{{2\hbar
c^2\varepsilon_0}Im\chi_m(\omega)}
\hat{C}_{\lambda'}(\textbf{k},\omega )\epsilon_{\lambda\lambda'}
e^{\imath{\bf k}\cdot\textbf{r}}.
\end{equation}
Considering (\ref{52}), (\ref{67}), and (\ref{62}), the ETCR between
$\hat {\bf M}(\bf r, \omega)$ and $\hat {\bf M}^\dagger(\bf r,
\omega')$ becomes

\begin{equation}\label{68}
[\hat{M}_{N\lambda}(\textbf{r},\omega),
\hat{M}_{N\lambda}^\dag(\textbf{r}',\omega')]={2\hbar
c^2\varepsilon_0}Im\chi_m(\omega)\delta_{\lambda
\lambda'}\delta(\textbf{r}-\textbf{r}')\delta(\omega-\omega').
\end{equation}
Relation (\ref{68}) satisfies dissipation$-$fluctuation theorem and
is the same as expression of noise operator in macroscopic method
\cite{12}. Therefor the supposed Lagrangian is equivalent to
macroscopic approach.
\section{Extension of the Huttner-Barnett model to a magnetodielectric medium}
In pervious section we showed that the HB model can be extended to a
magnetizable matter by changing the interaction between EM and the
matter fields. The extension of this model to a magnetodielectric
matter can be done by considering the following Lagrangian density
\begin{equation}\label{69}
{\cal L}={\cal L}_{em}+{\cal L}_{1mat}+{\cal L}_{1res}+{\cal
L}_{1int}+{\cal L}_{2mat}+{\cal L}_{2res}+{\cal L}_{2int},
\end{equation}
where ${\cal L}_{em}$ is defined in (\ref{11}), and
\begin{equation}\label{70}
{\cal L}_{1mat}=\frac{\rho}
{2}\dot{\textbf{X}}_1^2-\frac{\rho\omega_{0}^2}{2}\textbf{X}_1^2,
\end{equation}
\begin{equation}\label{71}
{\cal L}_{2mat}=\frac{\rho}
{2}\dot{\textbf{X}}_2^2-\frac{\rho\omega_{0}^2}{2}\textbf{X}_2^2,
\end{equation}
are the matter parts of Lagrangian, which are modeled by two
distinct harmonic oscillator fields $X_1$ and $X_2$ with the same
frequency $\omega_0$. In the following we call $X_1$ and $X_2$ the
polarization and magnetization fields respectively. It should be
noted that taking the same frequency for both polarization and
magnetization fields does not affect the result. The Lagrangians
describing the reservoir are defined by
\begin{equation}\label{72}
{\cal L}_{1res}=\int_0^\infty{d\omega}\left({\frac{\rho }{2}\dot{\bf
Y}_{1\omega}^2-\frac{{\rho\omega^2}}{2}{\bf Y}_{1\omega}^2} \right),
\end{equation}
\begin{equation}\label{73}
{\cal L}_{2res}=\int_0^\infty{d\omega}\left({\frac{\rho }{2}\dot{\bf
Y}_{2\omega}^2-\frac{{\rho\omega^2}}{2}{\bf Y}_{2\omega}^2} \right),
\end{equation}
and
\begin{equation}\label{74}
{\cal L}_{1int}=-\alpha_1{\bf{A}}\cdot\dot{\bf{X}}_1 -
\int_0^\infty{d\omega \nu_1 ( \omega
){\bf{X}}_1.\dot{{\textbf{Y}}}_{1\omega}},
\end{equation}
\begin{equation}\label{75}
{\cal L}_{2int}= \alpha_2 \nabla\times{\bf{A}}\cdot{\bf{X}}_2 -
\int_0^\infty{d\omega \nu_2( \omega
){\bf{X}}_2.\dot{{\textbf{Y}}}_{2\omega}},
\end{equation}
are the interaction parts. The coupling functions $\nu_1(\omega)$
and $\nu_2(\omega)$, satisfy the same assumptions we assumed for
$\nu(\omega)$ in (\ref{14}).

The displacement and magnetic fields are defined respectively by
\begin{equation}\label{76}
{\bf D}({\bf r},t) = \varepsilon _0 {\bf E}({\bf r},t)+{\bf P},
({\bf r},t)
\end{equation}
and
\begin{equation}\label{77}
{\bf H}({\bf r},t)=\kappa_0{\bf B}({\bf r},t)-{\bf M}({\bf r},t),
\end{equation}
where we have defined
$\textbf{P}(\textbf{r},t)\equiv-\alpha_1\textbf{X}_1(\textbf{r},t)$
and
$\textbf{M}(\textbf{r},t)\equiv+\alpha_2\textbf{X}_2(\textbf{r},t)$.

Again we choose the Coulomb gauge, but since the medium is
polarizable the scaler potential does not vanish. Since $\dot{U}$
does not appear in the Lagrangian, $U$ is not a proper dynamical
variable and should be written in terms of the proper dynamical
variables $\textbf{A}$, $\textbf{X}$, and $\textbf{Y}_\omega$. This
can be done by going to reciprocal space and writing the Lagrangian
as
\begin{equation}\label{78}
L = \int'{d^3{\bf k}(\tilde{\cal
L}_{em}+\sum_{\texttt{i}=1,2}\tilde{\cal L}_{\texttt{i}mat}+ \tilde
{\cal L}_{\texttt{i}res}+\tilde{\cal L}_{\texttt{i}int})}.
\end{equation}
The Lagrangian densities in this space are obtained as
\begin{equation}\label{79}
\tilde{{\cal L}}_{em}=\varepsilon_0\tilde{\bf
E}^2-\kappa_0\tilde{\bf B}^2,
\end{equation}

\begin{equation}\label{80}
\tilde{{\cal L}}_{\texttt{i}mat}=\rho
\dot{\tilde{\textbf{X}}}_\texttt{i}^2-\rho \omega _0^2\tilde{\bf
X}_\texttt{i}^2,
\end{equation}

\begin{equation}\label{81}
\tilde{{\cal L}}_{\texttt{i}res}=\int_0^\infty{d\omega}(\rho{\dot
{\tilde{\textbf{Y}}}}_{\texttt{i}\omega}^2-{{\rho \omega ^2
}}\tilde{{\bf Y}}_{\texttt{i}\omega}^2)
\end{equation}
with $\texttt{i}=1,2$, and
\begin{equation}\label{82}
\tilde{\cal L}_{1int}= [-\alpha_1
\tilde{\textbf{{A}}}^*\cdot\dot{\tilde{\textbf{X}}} - \int_0^\infty
d\omega \nu ( \omega
)\tilde{\bf{X}}\cdot\dot{\tilde{\textbf{Y}}}_\omega^*]+c.c.,
\end{equation}

\begin{equation}\label{83}
\tilde{\cal L}_{2int}= \alpha_2
\textbf{k}\times\tilde{\bf{A}}^*\cdot\tilde{\bf{X}}_2-
\int_0^\infty{d\omega\nu(\omega
)\tilde{\bf{X}}_2\cdot\dot{\tilde{\textbf{Y}}}_{2\omega}^*+c.c.}   .
\end{equation}
 Using the Euler-Lagrange equation for
${\dot{\tilde{U}}}^*$ we find
\begin{equation}\label{84}
\tilde{U}({\bf k},t) = \imath\frac{{\alpha _1 }}{{\varepsilon _0
}}(\frac{{{\bf k}\cdot\tilde{\bf X}({\bf k},t)}}{{{\bf k}^2 }}).
\end{equation}
Now if we decompose the longitudinal and transverse parts of the
fields, then the total lagrangian can be written as the sum of two
independent transverse and longitudinal parts as
\begin{equation}\label{85}
L = L^\parallel+ L^\perp,
\end{equation}
where
\begin{equation}\label{86}
L^\perp= \int'{d^3{\bf k}\tilde{\cal
L}^\perp_{em}+\sum_{\texttt{i}=1,2}(\tilde{\cal
L}^\perp_{\texttt{i}mat}+ \tilde {\cal
L}^\perp_{\texttt{i}res}+\tilde{\cal L}^\perp_{\texttt{i}int})},
\end{equation}
and
\begin{equation}\label{87}
\tilde{{\cal L}}^\perp_{em}=\varepsilon_0({ \tilde{\dot{\bf A}}
}^2-c^2{\bf \tilde B}^2),
\end{equation}

\begin{equation}\label{88}
\tilde{{\cal L}}^\perp_{\texttt{i}mat}=\rho
\dot{\tilde{\textbf{X}}}_\texttt{i}^{\perp2}-\rho \omega
_0^{\perp2}{\bf \tilde X}_\texttt{i}^{\perp2},
\end{equation}

\begin{equation}\label{89}
\tilde{{\cal
L}}^\perp_{\texttt{i}res}=\int_0^\infty{d\omega}(\rho{\dot
{\tilde{\textbf{Y}}}}_{\texttt{i}\omega}^{2}-\frac{{\rho \omega ^2
}}{2}\tilde{{\bf Y}}_{\texttt{i}\omega}^{\perp2}),
\end{equation}

\begin{equation}\label{90}
\tilde{\cal L}^\perp_{1int}= -\alpha_1
\tilde{\textbf{A}}\cdot\dot{\tilde{\textbf{X}}}^\perp -
\int_0^\infty{d\omega \nu_1 ( \omega
)\tilde{\bf{X}}^\perp\cdot\dot{\tilde{\textbf{Y}}}_\omega^{\perp*}+c.c.},
\end{equation}

\begin{equation}\label{91}
\tilde{\cal L}^\perp_{2int}= \alpha_2
\textbf{k}\times\tilde{\bf{A}}\cdot\tilde{\bf{X}}^\perp_2-
\int_0^\infty{d\omega\nu_2(\omega
)\tilde{\bf{X}}^{\perp}_2\cdot\dot{\tilde{\textbf{Y}}}_{2\omega}^{*\perp}+c.c.}.
\end{equation}
As in the previous section, we put away the longitudinal part of
magnetization field. So the longitudinal part of the Lagrangian
consists of only the polarization and EM field. Using (\ref{84}),
(\ref{79}) and (\ref{80}) the longitudinal part of the Lagrangian in
terms of the polarization field can be written as
\begin{equation}\label{92}
L^\parallel=\int '{d^3 {\textbf{k}}\tilde{\cal L}}_1 ^\parallel,
\end{equation}
where
\begin{eqnarray}\label{93}
\tilde{\cal L}_1^\parallel&=&\rho \tilde{\bf X}_1^{\parallel 2}  -
\rho \omega _L^2 \tilde{\bf X}_1^{\parallel 2}  + \int_0^{ + \infty
} {d\omega } (\rho \tilde{\bf Y}_{1\omega }^{\parallel 2} - \rho
\omega ^2 \tilde{\bf Y}_{1\omega }^{\parallel 2} )\nonumber\\&-&
\int_0^{ + \infty } {d\omega } v_1 (\omega )(\tilde{\bf
X}_1^{\parallel *}\cdot\tilde{\bf Y}_{1\omega }^\parallel   +
\tilde{\bf X}_1^{\parallel *}\cdot\tilde{\bf Y}_{1\omega }^\parallel
).
\end{eqnarray}
The $\omega_L$ is the longitudinal frequency and is defined by
$\omega_L\equiv\sqrt{\omega_0^2+\omega_c^2}$ where
$\omega_c^2=\frac{\alpha_1^2}{\rho\epsilon_0}$. The link between the
transverse and longitudinal parts is given by the total electric
field, which is written as
\begin{equation}\label{94}
\tilde{\textbf{E}}({\textbf{k}},t) =  -
\dot{\tilde{\textbf{A}}}({\textbf{k}},t) + \frac{\alpha_1
}{{\varepsilon _0 }}\tilde{\textbf{X}}^\parallel({\textbf{k}},t).
\end{equation}
Using (\ref{94}) and the definition of the displacement field $\bf
D$ given by (\ref{76}), we recover the fact that the displacement
vector is a purely transverse field.

In this work, we are mainly interested in the transverse fields and
shall only present the detailed quantization of the transverse part
of the Lagrangian. In what follows we restrict ourselves to
transverse fields and omit the superscript $\perp$. We use the same
unit polarization vectors $\textbf{e}_\lambda(\textbf{k})$,
$\lambda=1,2$ an find the the conjugate variables from $\tilde{{\cal
L}}$ as
\begin{equation}\label{95}
-\varepsilon_0\tilde{E}_\lambda=\frac{\partial{\cal
L}}{\partial\dot{\tilde{A}}_\lambda^*}=\varepsilon_0\dot{\tilde{A}}_\lambda,
\end{equation}

\begin{equation}\label{96}
\tilde {P}_{1\lambda}=\frac{\partial{\cal L}}
{\partial{\dot{\tilde{X}}_{1\lambda}^*}}=
\rho\dot{\tilde{X}}_{1\lambda}-\alpha_1\tilde{A}_\lambda,
\end{equation}

\begin{equation}\label{97}
\tilde {P}_{2\lambda}=\frac{\partial{\cal L}}
{\partial{\dot{\tilde{X}}_{2\lambda}^*}}=
\rho\dot{\tilde{X}}_{2\lambda},
\end{equation}

\begin{equation}\label{98}
\tilde{Q}_{\texttt{i}\omega\lambda}= \frac{\partial{\cal
L}}{\partial \dot{\tilde{Y}}_{\texttt{i}\omega\lambda}^*}=\rho\dot
{\tilde{Y}}_{\texttt{i}\omega\lambda}-v_\texttt{i}(\omega)
\tilde{Y}_{\texttt{i}\omega\lambda}.
\end{equation}

For the particular type of the coupling between light and
polarization field (?), the conjugate of $\tilde{\textbf{A}}$ is the
transverse electric field $-\epsilon_0\tilde{\textbf{E}}$. A
canonical transformation leading to a $\tilde{\textbf{E}}\cdot
\tilde{\textbf{X}}$ type of coupling, gives the displacement field
$-\tilde{\textbf{D}}$ as the conjugate of $\tilde{\textbf{A}}$.
Naturally, these two possibilities lead to the same results. We
choose here the first possibility in order to keep as close as
possible to the classical theory, where $\bf E$ is usually
considered as the fundamental variable.

Following pervious section we can obtain the Hamiltonian from the
Lagrangian (\ref{86}) and the conjugate variables
(\ref{95})-(\ref{98}), as
\begin{equation}\label{99}
H=\int'{d^3{\bf k}(\tilde{{\cal
H}}_{em}+\sum_{\texttt{i}=1,2}\tilde{{\cal H}}_{\texttt{i}mat}+
\tilde{{\cal H}}_{\texttt{i}int})},
\end{equation}
where

\begin{equation}\label{100}
\tilde{\cal H}_{em}=\epsilon_0(\tilde{\textbf{E}})^2
+\epsilon_0\tilde{\omega}_\textbf{k}^2\tilde{\textbf{A}}^2,
\end{equation}
is the electromagnetic energy density and
$\tilde{\omega}_\textbf{k}$ is defined by
$\tilde{\omega}_\textbf{k}^2\equiv c^2(k^2+k^2_c)$ with
$k_c\equiv\frac{\omega_c}{c}=\sqrt{\frac{\alpha_1^2}{\rho
c^2\epsilon_0}}$, the Hamiltonian densities

\begin{eqnarray}\label{101}
\tilde{\cal H} _{\texttt{i}mat}&=&
\frac{\tilde{\textbf{P}}_\texttt{i}^{2\perp}}{\rho
}\mathop{+\rho\tilde{\omega}_0^2\tilde{\textbf{X}}_\texttt{i}^{2\perp}+}\int_0^\infty
d\omega (\frac{{\tilde{\textbf{Q}}_{\texttt{i}\omega}^\perp}}{{\rho
}}^2+{\rho}\omega^2 \tilde{\textbf{Y}}_{\texttt{i}\omega}
^{\perp2})\nonumber\\&+&\int_0^\infty {d\omega }
(\frac{{V_\texttt{i}(\omega)}}
{\rho}\tilde{\textbf{X}}_\texttt{i}^{\perp*}\cdot\tilde{\textbf{Q}}
_{\texttt{i}\omega}^\perp)+ c.c.,
\end{eqnarray}
are the energy densities of the matter fields, where
$V_\texttt{i}(\omega),\,\,(i=1,2)$, are defined in (\ref{45}), and

\begin{equation}\label{102}
\tilde{\cal H}_{1in}=\frac{\alpha_1}{\rho_1}(\tilde{\bf A}^*
\cdot\tilde{\bf P}_1+c.c.),
\end{equation}

\begin{equation}\label{103}
\tilde{\cal H}_{2in}=-\frac{\alpha_2}{\rho_2}({\bf
k}\times\tilde{\bf A}^*\cdot\tilde{\bf X}_2+c.c.),
\end{equation}
are the interaction energies between the EM field and the matter
fields.

As usual, we demand ETCR between the variables and their conjugates
and introduce the annihilation operators as
\begin{equation}\label{104}
\hat{a}_\lambda(\textbf{k},t) = \sqrt {\frac{\epsilon_0}{{2\hbar
\tilde{k}c}}}(\tilde{k}c
\hat{\tilde{A}}_\lambda(\textbf{k},t)-{\imath}\hat
{\tilde{E}}_{\lambda}^{*}(\textbf{k},t)),
\end{equation}

\begin{equation}\label{105}
\hat{b}_{\texttt{i}\lambda}(\textbf{k},t) = \sqrt {\frac{\rho
}{{2\hbar \tilde{\omega}_0}}}(\tilde{\omega}_0
\hat{\tilde{X}}_{\texttt{i}\lambda}(\textbf{k},t)+\frac{\imath}{\rho}\hat
{\tilde{P}}_{\texttt{i}\lambda}^{*}(\textbf{k},t)),
\end{equation}

\begin{equation}\label{106}
\hat{b}_{\texttt{i}\lambda}(\textbf{k},,\omega,t)=\sqrt{\frac{\rho}{{2\hbar
\omega}}}
(-\imath\omega\hat{\tilde{Y}}_{\texttt{i}\omega\lambda}(\textbf{k},t)+
\frac{1}{\rho}\hat{\tilde{Q}}_{\texttt{i}\omega\lambda}^{*}(\textbf{k},t)),
\end{equation}
which satisfy the standard bosonic commutation relations.

The Hamiltonian (\ref{99}) can be written in terms of the
annihilation and creation operators defined in
Eqs.(\ref{107})$-$(\ref{109}). Since the structure of the matter
fields in the Hamiltonian (\ref{101}) are the same as $H_{mat}$ in
(\ref{28}), so the Hamiltonian of the matter can be diagonalized
using the same method. The total Hamiltonian in terms of
eigenoperators of the matter and reservoir fields can be written
as

\begin{eqnarray}\label{107}
\hat{H} &=& \int {d^3{\bf k}}\{ \hbar\tilde{\omega}_{\bf k}
\hat{a}_\lambda ^\dag ({\bf k})\hat{a}_\lambda({\bf
k})+\int_0^\infty{d\omega\hbar\omega}
\sum_{\texttt{i}}\hat{B}_{\texttt{i}\lambda }^\dag ({\bf k},\omega
)\hat{B}_{\texttt{i}\lambda }({\bf k},\omega )\nonumber\\ &+&
\frac{\hbar}{2}\Lambda _1({k})\int_0^\infty{d\omega}\{ g_1(\omega
)\hat{B}_{1\lambda }^\dag ({\bf k},\omega)[\hat{a}_\lambda({\bf
k})+\hat{a}_\lambda ^\dag(-{\bf
k})]+ H.c.\} \nonumber\\
\nonumber\\ &-& \frac{\hbar}{2}\Lambda _2({
k})\int_0^\infty{d\omega}\sum_{\lambda'}\{ g_2(\omega
)\hat{B}_{2\lambda' }^\dag ({\bf k},\omega)[\hat{a}_\lambda({\bf
k})+\hat{a}_\lambda
^\dag(-{\bf k})]\epsilon_{\lambda\lambda'}+ H.c.\} \},\nonumber\\
\end{eqnarray}
where the annihilation operators of the polarization and
magnetization fields are
\begin{equation}\label{108}
\hat{b}_{\texttt{i}\lambda}(\textbf{k},t)=\int_0^\infty{d\omega[\alpha_0
(\omega)}\hat{B}_{\texttt{i}\lambda}^\dag(\omega,\textbf{k},t)-\beta_0(\omega)
\hat{B}_{\texttt{i}\lambda}(\omega,\textbf{k},t)],
\end{equation}
where $\Lambda_1({k})\equiv\sqrt{\frac{\tilde{\omega}_0c
k_c^2\alpha_1^2}{\tilde{k}}}$ , $ \Lambda_2 (k) = \sqrt
{\frac{\alpha_2^2k^2}{{\varepsilon _0 \rho
\tilde{\omega}_\textbf{k}\tilde{\omega}_0}}}$,
$g_1(\omega)=\imath(\alpha_{01}(\omega)+\beta_{01}(\omega))$,
$g_2(\omega)=(\alpha_{02}^*(\omega)-\beta^*_{02}(\omega))$ and the
$\bf k$ integration has been extended to full reciprocal space.

In Eq.(\ref{107}), EM field is coupled with two distinct reservoirs
and is different from the usual Huttner model which only contains
one reservoir, so we represent here the details of the
diagonalization process of the Hamiltonian (\ref{107}).

The diagonalization of $\hat{H}$ can be achieved by introducing the
operators $\hat{C}(\bf k, \omega)$ as
\begin{eqnarray}\label{109}
\hat C_\lambda({\bf k},\omega)&=&\alpha_0(k,\omega)\hat
a_\lambda({\bf k})+\beta_0 (k,\omega)\hat a_\lambda^\dag({\bf
k})\nonumber\\&+&\int_0^\infty{d\omega'}[\alpha_1(k,\omega,\omega
')\hat B_1({\bf k},\omega')+\beta _1(k,\omega,\omega ')\hat B_1^\dag
({\bf k},\omega ')]\nonumber\\&+&\int_0^\infty{d\omega
'}\sum_{\lambda'}[\alpha _2 (k,\omega,\omega ')\hat
B_{2\lambda'}({\bf k},\omega ')+\beta_2(k,\omega ,\omega ')\hat
B_{2\lambda'}^\dag ({\bf k},\omega
')]\epsilon_{\lambda,\lambda'},\nonumber\\
\end{eqnarray}
where the coefficients are chosen such that the operators
$\hat{C}({\bf k},\omega )$ satisfy the eigenoperator equation
\begin{equation}\label{110}
[\hat{C}({\bf k},\omega ),\hat{H}] = \hbar \omega \hat{C}({\bf
k},\omega ).
\end{equation}

This equation, together with the expansion of the Hamiltonian
(\ref{107}) and the definition of $\hat{C}(\bf k,\omega)$ in
(\ref{109}), lead to the following linear equations between the
coefficients
\begin{eqnarray}\label{111}
\alpha_0({ k},\omega)\omega&=&\alpha_0({k},\omega)\omega_{\bf
k}\nonumber\\&+&\sum_{\texttt{i}=1,2}
\frac{1}{2}\int_0^\infty{d\omega'}[\alpha_1({
k},\omega,\omega')V_\texttt{i} ({k},\omega ')-\beta_\texttt{i}^*({
k},\omega ,\omega')V_\texttt{i}({
k},\omega ')],\nonumber\\
\end{eqnarray}
\begin{eqnarray}\label{112}
\beta_0({k},\omega)\omega&=&-\beta_0({k},\omega)\omega_{\bf
k}\nonumber\\&+&\sum_{\texttt{i}=1,2}
\frac{1}{2}\int_0^\infty{d\omega'}[\alpha_\texttt{i}(
{k},\omega,\omega')V_\texttt{i} ({k},\omega
')-\beta_\texttt{i}({\bf k},\omega
,\omega')V_\texttt{i}({k},\omega
')], \nonumber\\
\end{eqnarray}

\begin{equation}\label{113}
 \alpha_\texttt{i}({k},\omega,\omega')\omega=
\frac{1}{2}[\alpha _0 ({k},\omega )-\beta_0({
k},\omega)]V_\texttt{i}^*({k},\omega')+ \alpha_\texttt{i}({
k},\omega,\omega')\omega',
\end{equation}

\begin{equation}\label{114}
 \beta _\texttt{i}({k},\omega,\omega')\omega=
\frac{1}{2}[\alpha _0 ({k},\omega)-\beta_0({
k},\omega)]V_\texttt{i}^*({k},\omega ')-\beta_\texttt{i}({k},\omega
,\omega ')\omega ',
\end{equation}
where
$V_\texttt{i}(k,\omega)=\Lambda_\texttt{i}(k)g_\texttt{i}(\omega)$.

These set of equations can be easily solved to obtain
$\beta_0({k},\omega)$, $\alpha_\texttt{i}({k},\omega,\omega')$ and
$\beta _\texttt{i}({k},\omega,\omega')$ in terms of $\alpha_0({\bf
k},\omega)$. Subtracting (\ref{112}) from (\ref{111}) we obtain
\begin{equation}\label{115}
\beta _{0} ({k},\omega) = \frac{\omega-\tilde{\omega}_{\bf
k}}{\omega+\tilde{\omega}_{\bf k} }\alpha_{0}({k},\omega).
\end{equation}
We now replace (\ref{115}) for $\beta_0({\bf k},\omega)$ in
(\ref{113}) and (\ref{114}), and find
\begin{equation}\label{116}
\alpha_\texttt{i}({
k},\omega,\omega')=[P(\frac{1}{\omega-\omega'})+y_\texttt{i} ({
k},\omega )]V_\texttt{i}^*({k},\omega')\frac{{\tilde{\omega}_{
\textbf{k}}}}{\omega+\tilde{\omega}_{\textbf{k}}
}\alpha_0(\omega),
\end{equation}

\begin{equation}\label{117}
\beta_\texttt{i}({
k},\omega,\omega')=[\frac{1}{\omega+\omega'}]V_\texttt{i}({
k},\omega')\frac{{\tilde{\omega}_{\bf
k}}}{\omega+\tilde{\omega}_{\bf k} }\alpha_0({k},\omega),
\end{equation}
where P means the Cauchy principal value. The relation between
functions $y_1({ k},\omega)$ and $y_2({k},\omega)$ can be obtained
by substituting the expressions for $\alpha_i({k},\omega,\omega')$
and $\beta_i({k},\omega,\omega')$ in (\ref{116}) and (\ref{117})
into (\ref{111}). Using the definitions of $V_1^2({ k},\omega)$ and
$V_2^2({ k},\omega$ in (\ref{107}) it is easy to show that they are
odd functions of frequency $\omega$. We use this fact to extend the
integral in the negative frequency region and obtain the relation
between functions $y_1({ k},\omega)$ and $y_2({k},\omega)$ as

\begin{eqnarray}\label{118}
V_1^2({k},\omega) y_1({k},\omega )+V_2^2 ({k},\omega)y_2 ({
k},\omega )&=&\frac{{\omega ^2  - \tilde{\omega}_{\bf k}^2
}}{{\tilde{\omega} _{\bf k} }} + \frac{1}{2}P\int_{ - \infty }^{ +
\infty } {d\omega '}\frac{V_1^2 ({k},\omega ')}{\omega ' - \omega
}\nonumber\\&+& \frac{1}{2}P\int_{ - \infty }^{ + \infty }
{d\omega '} \frac{V_2^2 ({
k},\omega ')}{\omega '-\omega}.\nonumber\\
\end{eqnarray}

In order to calculate $\alpha_0({k},\omega)$, we impose the standard
commutation relation on $\hat C (\bf{k},\omega)$
\begin{equation}\label{119}
[\hat{C}({\bf k},\omega),\hat{C}^\dag({\bf
k}',\omega)]=\delta(\omega-\omega')\delta({\bf k}-{\bf k}').
\end{equation}
Using the expression for $\hat C(\bf k,\omega)$ given by
(\ref{109}) and the set of equations $\alpha_0(k,\omega)$
(\ref{115})$-$(\ref{117}), we can find $\alpha_0(k,\omega)$ (up to
a phase factor) in terms of the  $y_1({ k},\omega)$ and $y_2({
k},\omega)$. By taking a suitable phase factor and doing some
routine but tedious calculations we find the following expression
for $\alpha _0 ({k},\omega )$

\begin{eqnarray}\label{120}
\alpha _0 ({k},\omega )=\frac{\omega+\tilde{\omega} _{\bf
k}}{\tilde{\omega} _{k}}\{\frac{1}{(y_1 ({k},\omega ) + \imath\pi
)^2 V_1^2({k},\omega) + (y_2 ({k},\omega ) + \imath\pi )^2 V_2^2({
k},\omega)}\}^{\frac{1}{2}}.\nonumber\\
\end{eqnarray}

 From the Eqs.(\ref{109}) and (\ref{111})$-$(\ref{114}), we can
 obtain two independent sets of operators, $\hat{C}$ and $\hat{C}'$
 which satisfy the following commutation relation
 \begin{equation}\label{121}
[\hat{C}(k,\omega),\hat{C}'^{\dag}(k,\omega)]=0.
\end{equation}
For obtaining these operators, we first choose $y_1(k,\omega)$ and
$y_2(k,\omega)$ such that they satisfy (\ref{118}) and then using
(\ref{120}), we find $\hat{C}$ and $\hat{C}^{\dag}$.  For obtaining
$\hat{C}'$ and $\hat{C}'^{\dag}$, we should choose $y_1'$ and $y_2'$
such that they satisfy (\ref{118}) and also the derived operators
$\hat{C}'$ and $\hat{C}'^{\dag}$ from them should satisfy
(\ref{121}). So, for defining $y_1'$ and $y_2'$ there are two
equations and accordingly they can be determined uniquely. operator
$\hat C$.

Since in Eq.(\ref{118}), $y_1(k,\omega)$ and $y_2(k,\omega)$ can not
be determined uniquely, so we have a freedom in determining
operators $\hat C$, $\hat C'$. But we do not lose any generality by
taking a special solution since these operators are all equivalent
up to a Bogoliubov transformation.

To facilitate the calculations, we choose
 $y_1(k,\omega)$ and $y_2(k,\omega)$ such that they satisfy in
 (\ref{118}) and the following relation
\begin{equation}\label{122}
V_1({k},\omega) [y_1 ({k},\omega ) - \imath\pi ] =  +
V_2({k},\omega) [y_2 ({k},\omega ) - \imath\pi ].
\end{equation}
Therefore $y_1(k,\omega)$ is obtained as

\begin{eqnarray}\label{123}
y_1 ({k},\omega )&=& \frac{1}{{V_1^2 ({k},\omega) + V_1({ k},\omega)
V_2({k},\omega) }}\{ \frac{{\omega ^2  - \tilde{\omega} _{\bf k}^2
}}{{\tilde{\omega} _{\bf k} }}+ \frac{1}{2}\int_{ - \infty }^\infty
{d\omega '} \frac{{V_1 ({k},\omega)}}{{\omega ' - \omega
}}\nonumber\\ &+& \frac{1}{2}\int_{ - \infty }^\infty {d\omega '}
\frac{{V_2 ({k},\omega)}}{{\omega ' - \omega }} - \imath\pi
V_2^2({k},\omega) (\frac{{V_1 ({ k},\omega)}}{{V_2 ({k},\omega)}} -
1)\}.\nonumber\\
\end{eqnarray}
Using (\ref{120}) and (\ref{123}), we find $\alpha_0(k,\omega)$ as
\begin{equation}\label{124}
\alpha_0({k},\omega ) = \frac{\omega+\tilde{\omega}_{\bf k}
}{\sqrt 2}(\frac{V_1({k},\omega)+ V_2({k},\omega)}{\omega ^2 -
\tilde{\omega} _{\bf k}^2  + z_1 ({k},\omega)+z_2 ({k},\omega)}),
\end{equation}
where $z_\texttt{i}({
k},\omega)\equiv\frac{\tilde{\omega}_\textbf{k}}{2}\int_0^\infty
\frac{V_\texttt{i}^2({k},\omega)} {\omega-\omega'+i\epsilon}$. The
other set of operators, $\hat C'(\bf k, \omega)$, can be obtained
from
\begin{equation}\label{125} V_1 ({k},\omega)[y'_1
({k},\omega) - \imath\pi ] =  - V_2 ({ k},\omega)[y'_2 ({k},\omega)
- \imath\pi ].
\end{equation}
For $\alpha'_0(k,\omega)$ we have

\begin{equation}\label{126}
\alpha'_0({k},\omega ) = \frac{\omega + \tilde{\omega} _{\bf k}
}{\sqrt 2}(\frac{{V_1({k},\omega) -V_2({k},\omega) }}{{\omega ^2 -
\tilde{\omega} _{\bf k}^2  + z_1 ({k},\omega) + z_2 ({
k},\omega)}}).
\end{equation}
Eqs.(\ref{124}) and (\ref{126}) can be used to obtain $C$ and $C'$
in terms of $\hat a$, $\hat a^\dagger$, $\hat B_i$ and $\hat
B_i^\dagger$. From (\ref{110}), (\ref{119}) and (\ref{121}) we can
write the Hamiltonian (\ref{107}) as
\begin{equation}\label{127}
\hat{H}=\int d^3 {\bf k}\int_{0}^\infty\hbar\omega
[\hat{C}^\dag({\bf k},\omega)\hat{C}({\bf k},\omega) +
\hat{C}'^\dag({\bf k},\omega)\hat{C}'({\bf k},\omega)].
\end{equation}
Using the commutation relation (\ref{119}) and the commutation
relation between $\hat a$ and $\hat a^\dagger$, we can invert the
Eq.(\ref{119}) to write $\hat a$ and $\hat a^\dagger$ in terms of
$\hat C$, $\hat C^\dag$, $\hat C'$ and $\hat C'^\dag$ as
\begin{eqnarray}\label{128}
\hat{a}({\bf k})&=&\int_0^\infty d\omega\{\alpha^*_0 ({k},\omega
)\hat{C}({\bf k},\omega ) - \beta _0 ({k},\omega )\hat{C}^\dag
({\bf k},\omega )\nonumber\\&+& \alpha'^* _0 ({k},\omega
)\hat{C}'({\bf k},\omega)- \beta'_0 ({k},\omega )\hat{C}'^\dag
({\bf k},\omega )\}.
\end{eqnarray}

  Before writing the EM field in terms of $\hat{C}$ and $\hat{C}^\dag$, for later
simplification we use a Bogoliubov transformation as

\begin{equation}\label{129}
\hat{K}_e ({\bf k},\omega ) = \frac{{\hat{C}({\bf k},\omega ) +
\hat{C}'({\bf k},\omega )}}{{\sqrt 2 }},
\end{equation}

\begin{equation}\label{130}
\hat{K}_m({\bf k},\omega ) = \frac{{\hat{C}({\bf k},\omega ) -
\hat{C}'({\bf k},\omega )}}{{\sqrt 2 }}.
\end{equation}
 Using (\ref{59}) and (\ref{128}), $\hat{\bf A}$ can be obtained in terms
 of eigenoperators of the Hamiltonian, as
\begin{eqnarray}\label{131}
\hat{{\bf A}}({\bf r},t) &=&
\frac{\imath}{(2\pi)^\frac{3}{2}}\sqrt\frac{\hbar}{2\varepsilon_0}\int
d^3 {\bf k}\int_{0}^{\infty}d\omega \{[\frac{\omega{f_1 (\omega
)}\hat{K}_e(\textbf{k},\omega)}{{\omega _{\bf k} ^2 (1 - \chi _m
(\omega )) - \omega ^2 (1 + \chi _e (\omega
))}}\nonumber\\&+&\frac{\omega_\textbf{k}{f_2 (\omega
)}\hat{K}_m(\textbf{k},\omega)}{\omega _{\bf k}^2 (1 -\chi _m
(\omega )) - \omega ^2 (1 + \chi _e (\omega
))}]e^{\imath(\textbf{k}\cdot \textbf{r}-\omega t)}-H.c.\},\nonumber\\
\end{eqnarray}
where
\begin{equation}\label{132}
\chi_e (\omega ) \equiv\frac{1}{2}\int_{ - \infty }^{+\infty}d\omega
'\frac{f_1^2(\omega ')}{{\omega - \omega ' - \imath\varepsilon
}}=\frac{1}{2} P\int_0^\infty d\omega'\frac{f_1^2(\omega ')}{{\omega
- \omega '
 }}+\frac{1}{2}\imath\pi |f_1(\omega )|^2,
\end{equation}
and
\begin{equation}\label{133}
\chi_m (\omega )\equiv \frac{1}{2}\int_{ - \infty }^{+\infty}d\omega
'\frac{f^2_2(\omega ')}{{\omega - \omega ' - \imath\varepsilon
}}=\frac{1}{2}P\int_0^\infty d\omega'\frac{f_2^2(\omega ')}{{\omega
- \omega '
 }}+\frac{1}{2}\imath\pi |f_2(\omega )|^2,
\end{equation}
and $f_1(\omega)\equiv{\sqrt{\tilde{\omega}_0c^2
k_c^2}}\frac{\alpha_1g_1(\omega)}{\omega}
 $ and $f_2(\omega)\equiv\frac{\alpha_2g_2(\omega)}
 {\sqrt{c^2\tilde{\omega}_0\varepsilon_0\rho}}$.

 Now the process of calculating $\hat{\bf M}$ in the previous
 section can be repeated for $\hat{{\bf P}}$ and $\hat{{\bf M}}$
 ($\hat{{\bf P}}$ and $\hat{{\bf M}}$ are defined in (\ref{76}) and (\ref{77})
 respectively). We find
\begin{equation}\label{134}
{\hat{\textbf{P}}}({\bf r},t) = \int_0^\infty  {} d\omega
\{[\varepsilon_0\chi _e (\omega ){\bf \hat{\textbf{E}}}({\bf
r},\omega ) +{\hat{\textbf{P}}}_N ({\bf r},\omega )]
e^{-\imath\omega t}+H.c.\},
\end{equation}
and
\begin{equation}\label{135}
{\hat{\textbf{M}}}({\bf r},t) = \int_0^\infty d\omega
\{[\kappa_0\chi _m (\omega )\nabla\times{\hat{\textbf{A}}}({\bf
r},\omega ) + {\hat{\textbf{M}}}_N ({\bf r},\omega )]e^{ -
\imath\omega t}+H.c.\},
\end{equation}
where $\hat{{\bf P}}_N(\bf r ,\omega)$ and $\hat{{\bf M}}_N(\bf r
,\omega)$ are
\begin{equation}\label{136}
 \hat{P}_{N\lambda}({\textbf{r}},\omega)=\int d^3 \textbf{k}
 \sqrt{2\hbar\varepsilon_0 Im\chi_e}\hat{K}_{e\lambda}({\bf
 k},\omega)e^{\imath\textbf{k}\cdot\textbf{r}},
 \end{equation}

\begin{equation}\label{137}
 \hat{M}_{N\lambda}({\textbf{r}},\omega)=\int d^3\textbf{k}\sum_{\lambda'=1,2}
 \sqrt{2\hbar\varepsilon_0 c^2 Im\chi_m}
 \hat{K}_{m\lambda'}({\bf k},\omega)e^{\imath\textbf{k}\cdot\textbf{r}}
 \epsilon_{\lambda\lambda'}.
\end{equation}

 By comparing (\ref{134}), (\ref{135}) and (\ref{8}), (\ref{9}) we find
 that $\chi_e$ and $\chi_m$ are
 electric and magnetic susceptibilities. As in the preview section, the
 commutation relation between $\hat{\textbf{P}}_N(\bf r,\omega)$ and
  $\hat{\textbf{P}}^\dagger_N(\bf r,\omega)$ and $\hat{\textbf{M}}_N(\bf r,\omega)$
  and $\hat{\textbf{M}}^\dagger_N(\bf r,\omega)$
  can be calculated. The results are compatible with
 the dissipation-fluctuation theorem and coincide with macroscopic results.

Using relation (\ref{131}), (\ref{132}) and Eq.(\ref{133}), we find
\begin{eqnarray}\label{138}
\hat{{\bf A}}({\bf r},t) &=&-\imath(\frac{1}{8\pi^4\varepsilon_0})
\int d^3 {\bf k}\int_{0}^{\infty}d\omega
\{[\frac{\omega\sqrt{Im\chi_e (\omega
)}\hat{K}_e(\textbf{k},\omega)}{{\omega _{{\bf k} }^2 (1 - \chi _m
(\omega )) - \omega ^2 (1 + \chi _e (\omega
))}}\nonumber\\&+&\frac{\omega_\textbf{k}\sqrt{Im\chi_m (\omega
)}\hat{K}_m(\textbf{k},\omega)}{\omega _{\bf k}^2 (1 - \chi _m
(\omega )) - \omega ^2 (1 + \chi _e (\omega
))}]e^{-\imath(\omega t-\textbf{k}\cdot \textbf{r})}-H.c.\}.\nonumber\\
\end{eqnarray}

The relations (\ref{138}), (\ref{132}) and (\ref{133}), and the
commutation relations between the noise operators are exactly the
same as the results obtained from the macroscopic method
\cite{12,15}. So, these two methods are equivalent.

\section{conclusion}
The Huttner-Barnett model has been extended to a magnetodielectric
medium. The results obtained in the present model are equivalent
with those obtained in the phenomenological models. The explicit
form of the noise operators have been obtained. Based on the
results obtained here, the Lagrangian introduced in the present
work can be used as a microscopic model for canonical quantization
of the electromagnetic field in a magnetodielectric medium.

\end{document}